\documentclass[12pt]{article}
\usepackage[T2A]{fontenc}
\usepackage[cp1251]{inputenc}
\usepackage{graphicx}
\newcommand{\be}{\begin{equation}}
\newcommand{\ee}{\end{equation}}
\begin{document}
\large
\baselineskip=20pt
\centerline{Printed in: \bf Astronomy Report, 2003, v.47, p.197}
\bigskip

\centerline{\bf
Supersoft
X-ray
Sources.
Basic
Parameters.}
\normalsize

\begin{center}
V.F.Suleimanov$^1$,
A.
A.
Ibragimov$^1$,\\
{\it 1 - Kazan State
University,
Kazan,
Tatarstan,
Russia}\\

Received
March
5,
2002;
in
final
form,
October
10,
2002
\end{center}

\bigskip

\centerline {\bf Abstract}

The
parameters
of
ten
supersoft
X-ray
sources
(RX J0439.8-6809,
RX J0513.9-6951,
RX J0527.8-6954,
CAL 87,
CAL 83,
1E 0035.4-7230,
RX J0048.4-7332,
1E 0056.8-7154,
RX J0019.8 +2156,
RX J0925.7-4758)
observed
by
ROSAT
obtained
using
blanketing
approximations
and
LTE
model
atmospheres
are
analyzed.
The
consistency
of
the
resulting
parameters
with a
model
with
stable/recurrent
burning
on
the
surface
of
the
white
dwarf
is
studied.
The
luminosity
and
sizes
of
seven
of
the sources
are in good agreement with this model. The masses of the white
dwarfs in these
sources
are
estimated.
A formula that
can be used to estimate the masses of white dwarfs in classical supersoft
sources based on their effective temperatures is presented.
\bigskip

\centerline{\bf 1. Introduction}
\bigskip

Supersoft
X-ray
sources
are a
class
of
X-ray
object
with
very
soft
spectra
(effective
temperatures
of
the
order
of
10 - 80
eV)
(Kahabka \& van den Heuvel 1997).
It
is
known
that
supersoft
sources
do
not
form a
uniform
class.
Some
are
close
binary
systems
with
periods
from
tenths
of
days
(0$^d$.17
for
1E0035 - typical
of
the
periods
of
cataclysmic
variables)
to
several
days
(1$^d$.04
for
CAL83).

The
widely
adopted
classical
model
for
these
sources
put
forth
by
van
den
Heuvel
et al. (1992, vdH92)
proposed
that
they
are
close
binary
systems
with a
white
dwarf
and
subgiant,
in
which
material
from
the
donor
is
accreting
onto
the
white
dwarf
on
the
thermal
time
scale
at a
rate
on
the
order
of
10$^{-7}$
M/yr.
At
such
accretion
rates
onto
the
white-dwarf
surface, a
regime
of
stable
thermonuclear
hydrogen
burning
is
realized,
without
the
substantial
increase
in
the
radius
of
the
white
dwarf
predicted
theoretically
by
Paczynski
\&
Zytkow (1978),
Iben
(1982),
Nomoto
(1982),
Fujimoto
(1982),
Iben \&
Tutukov
(1996).
However,
this
model
also
encounters
certain
difficulties.
In
particular,
the
radial
velocities
of
the
source
CAL87
imply a
mass
for
the
secondary
that
is
half
that
of
the
white
dwarf,
in
contradiction
with
the
expectations
of
the
model
(Cowley et al. 1998).

Symbiotic
stars
with
periods
of
hundreds
of
days
($\sim$ 550d
for
AG
Dra)
are
also
found
among
supersoft
sources.
It
is
thought
that,
in
these
systems,
there
is
accretion
from
the
wind
of a
giant
donor,
with
subsequent
burning
on
the
white-dwarf
surface.
It
is
likely
that
recurrent
novae
(such
as U
Sco
(Kahabka et al. 1999a))
also
undergo a
supersoft
stage.
The
source
RX J0439
in
the
Large
Magellanic
Cloud
has
not
yet
been
demonstrated
to
be a
binary.
There
are
no
signs
of
an
accretion
disk
in
its
spectrum,
so
that
it
has
been
classified
either
as a
single
star
at a
very
late
stage
of
evolution
or
as a
system
of
two
degenerate
dwarfs
(Gansicke et al. 2000).

In
all
cases,
the
presence
of
soft
X-ray
emission
is
associated with thermonuclear burning on the surface
of a
white
dwarf
or a
naked
hydrogen
or
helium
core
of a
star
in a
late
stage
of
evolution.

In
the
first
part
of
our
current
study
(Ibragimov et al. 2003),
we
approximated
ROSAT
spectra
using
blanketed
LTE
model atmospheres. Here, we analyze the consistency
of
those
results
with
the
theory
of
stable/recurrent
burning
on
the
surface
of a
white
dwarf,
estimate a
number
of
physical
parameters
of
the
sources,
and
analyze
their
mass–effective
temperature
relation.
The
full
designations
of
the
sources
are
presented
in
Table 1.
We
will
use
shortened
designations
consisting
only
of
the
first
several
symbols.
\bigskip

\centerline{\bf 2. Position of sources in the Hertzsprung -
Russel diagram}
\bigskip

 In previous paper (Ibragimov et al. 2003), we approximated archival
ROSAT
observations
using
blanketed
LTE
model
atmospheres.
For
most
sources,
we
obtained
four
different
approximations:
for
the
smallest
and
largest
possible
values
of
$\log~g$
and
with
the
column
density
of
interstellar
hydrogen either treated as a free parameteror
fixed at the
Galactic
value
(i.e.,
the
value
for
the
given
direction
in
the
Galaxy).
The
resulting
source
parameters
are
presented
in
Tables
3 - 6
of
Ibragimov et al. (2003):
the
column
density
of interstellar hydrogen
$N_H$, effective temperature
$T_{\rm eff}$,
logarithm
of
the
gravitational
acceleration
$\log~g$, and
normalization
factor
$R^2/d^2$.
Knowing
the
source
parameters
and
distances $d$, we
can
find
their
luminosities
and
sizes:
\be
  L= 4\pi \sigma T^4_{\rm eff} \frac{R^2}{d^2} d^2, ~~~~~~~~~~~~
R = \sqrt{\frac{R^2}{d^2}} d.
\ee

We
took
the
distances
to
the
Large
and
Small
Magellanic
Clouds
to
be
50
and
60
kpc,
respectively.
We
calculated
the
luminosities
and
sizes
of
sources
in
the
Galaxy
assuming $d$ =
2
kpc.

The
positions
of
sources
in
the
Hertzsprung -
Russell
(HR)
diagram
are
shown
in
Fig.
1,
which
also
depicts
theoretical
curves
for
white
dwarfs
of
various
masses
with
hydrogen
burning
on
their
surfaces.
It
is
known
(Iben \& Tutukov 1996)
that
the
maximum
luminosity
of
such a
white
dwarf
depends
substantially
on
its
internal
temperature,
especially
in
the
case
of
low-mass
white
dwarfs.
The
luminosity
of a
hot
white
dwarf
whose
temperature
is
comparable
to
the
temperature
in a
shell
with
thermonuclear
burning,
\be
\frac{L}{L_{\odot}} = 60000 \left(\frac{M}{M_{\odot}}-0.52\right),
\ee
is
substantially
lower
than
the
luminosity
reached
in
the
case
of
thermonuclear
burning
on
the
surface
of a
cooled
white
dwarf,
\be
\frac{L}{L_{\odot}} = 46000 \left(\frac{M}{M_{\odot}}-0.26\right).
\ee
Since
we
do
not
know
the
temperatures
of
the
white
dwarfs
in
the
studied
sources a
priori, we compared
their
positions
in
the
HR
diagram
with
evolutionary
tracks
for
novae
in
the
declining
brightness
phase,
which
corresponds
to
burning
on
the
surfaces
of
cool
white
dwarfs
(Kato \& Hachisu 1994)
(Fig.
1a),
and
with
tracks
for
burning
on
the
surfaces
of
hot
white
dwarfs,
above
the
edge
of
the
stable-burning
strip
(Iben \& Tutukov 1996)
(SBS; the
zone
in
which
there
can
be
continuous
burning;
Fig.
1b).
The
horizontal
part
of
the
tracks
reflects
variations
in
the
photospheric
radius
and
effective
temperature
of
the
white
dwarf
with
the
luminosity
remaining
nearly
constant
at
its
maximum
value,
while
the
sloped
part
of
the
tracks
(Fig.
1a)
reflects
variations
in
the
luminosity
as
the
temperature
of
the
white
dwarfs
decreases,
with
the
radius
remaining
constant.

Four
sources
(RX
J0925,
CAL
87,
RX
J0513,
and
1E
0056)
are
located
near
turning
points
of
the
evolutionary
curves
for
both
cool
and
hot
white
dwarfs.
This
means
that
they
are
located
in
the
SBS,
and
that
their
luminosities
are
close
to
the
maximum
values
(without
appreciable
increases
in
the
white-
dwarf
radii).
Note
that
their
positions
correspond
to
curves
for
white
dwarfs
of
various
masses,
depending
on
their
internal
temperatures
(compare
Figs.
1a
and
1b).
Three
sources
(CAL
83,
RX
J0527,
and
1E
0035)
are
located
on
white-dwarf
cooling
curves
below
the
SBS.

The
positions
of
four
sources
(1E
0035,
1E
0056,
CAL
83,
and
RX
J0527)
are
shown
for
two
values
of
$\log~g$
(the
lower
left
point
corresponds
to
$\log~g$ =
9.5,
and
the
upper
right
point
to
$\log~g$ =
8.0).
Since
the
parameters
obtained
for
RX J0048,
RX J0439,
and
RX J0019
were
either
uncertain
or
do
not
agree
with
acceptable
parameters
for
white
dwarfs,
these
objects
are
shown
only
in
Fig.
1a
for
the
minimum
value
of
$\log~g$.
The
luminosity
of
RX J0019
is
very
uncertain,
but
this
source
may
be a
white
dwarf
located
in
the
SBS.
RX J0439
and
RX J0048
have
luminosities
higher
than
the
Eddington
luminosities
of
white
dwarfs
(a
lower
limit
for
the
latter
source
is
indicated).

The
parameters
obtained
in
(Ibragimov et al 2003)
show a
large
scatter
due
to
uncertainty
in
the
interstellar
absorption
(determined
by
$N_H$).
However,
independent
estimates
of
$N_H$
derived
from
ultraviolet
observations
are
available
for
some
of
the
sources,
which
testify
that
the
values
of
$N_H$
in
the
direction
of
the
sources
in
the
Magellanic
Clouds
differ
little
from
the
Galactic
value.
Therefore,
we
used
the
parameters
for
these
sources
obtained
for
the
Galactic
value
of
$N_H$
(when
this
was
possible).
The
values
of
the
approximation
parameters
adopted
to
obtain
the
source
characteristics
are
presented
in
Table
1. A
basis
for
the
choice
of
one
or
another
approximation
for
each
source
is
given
below.
\bigskip

\centerline{\bf 3. Estimates of the source masses and sizes}
\bigskip

Figure 2
shows
the
positions
of
the
sources
in
the
$R$ - $\log~g$
plane.
The
bold
line
presents
the
theoretical
dependence
for
cool
white
dwarfs,
while
the
dashed
line
presents
the
same
dependence
for
the
case
when
the
white-dwarf
radius
is
twice
the
theoretical
value
for a
cool
white
dwarf.
The
circles
in
Fig.
2a
show
the
positions
of
six
sources:
three
with
fixed
values
of
$\log~g$
(CAL
87,
RX
J0925,
and
RX
J0513)
and
three
whose
sizes
are
very
uncertain
(RX
J0019)
or
are
not
consistent
with
the
sizes
expected
for
white
dwarfs
(RX
J0439
and
RX
J0048);
only
lower
limits
to
the
sizes
of
these
last
sources
are
indicated.
The
circles
in
Fig.
2b
show
the
positions
of
the
remaining
four
sources
for
the
extreme
values
of
$\log~g$.

Let
us
analyze
the
positions
of
the
sources
in
the
HR
diagram
(Fig.
1)
and
in
the
$R$ -
$\log~g$
plane
(Fig.
2)
taking
into
account
available
observational
information,
with
the
aim
of
estimating
the
masses
of
the
white
dwarfs
in
these
objects.
The
mass
can
be
derived
from
the
source
size
as
follows.
We
first
determine
the
regions
of
$\log~g$
values
in
Fig. 2
corresponding
to
the
sizes
of
the
sources,
to
the
theoretical
sizes
of
cool
white
dwarfs
(solid
line),
and
to
sizes
exceeding
these
theoretical
sizes
by a
factor
of
two
(dashed
line;
we
will
call
this
the
"2R
case").
In
this
way,
we
find
the
$\log~g$
values
for
which
the
theoretical
curves
correspond
to
the
source
sizes
derived
from
observations.
We
can
then
find
the
mass
of
an
object
from
the
two
$M$ ---
$\log~g$
relations,
calculated
from
the
$M$ --- $R$
law
for
cool
white
dwarfs
(Popham \& Narayan 1995)
and
for
objects
with
the
same
masses
but
doubled
radii
(Fig.
3).

{\it RX
J0439.8-6809.}
The
size
and
luminosity
of
this
source
testify
that
it
is
not a
classical
supersoft
source.
In
our
analysis,
we
used
the
parameters
derived
for
$\log~g$
=7.5
and
the
Galactic
$N_H$
value,
since
$N_H$
values
derived
independently
using
Hubble
Space
Telescope
(HST)
data
(G\"ansicke et al. 2000)
demonstrate
the
soundness
of
this
approach.
Two
models
are
currently
being
considered
for
this
source
in
the
literature.
It
may
be a
system
of
two
degenerate
dwarfs,
one
of
which
overfills
its
Roche
lobe,
while
thermonuclear
burning
occurs
on
the
surface
of
the
other
(Iben \& Tutukov 1993),
or
it
may
be a
star
such
as
PG 1159,
which
is
the
virtually
naked
core
of a
star
in a
late
stage
of
its
evolution
(G\"ansike et al. 2000).
In
either
case,
the
spectrum
of
this
source
cannot
be
modeled
using
hydrostataic
model
atmospheres
with
solar
chemical
composition,
and
the
parameter
values
we
have
obtained
should
be
treated
as
very
approximate.
It
is
not
possible
to
derive
its
mass
from
our
data.

{\it RX
J0513.9 - 6951.}
This
is a
recurrent
binary
whose
parameters
are
in
agreement
with
the
vdH92
model.
The
value
of
$N_H$
has
been
derived
using
HST
data
(G\"ansicke et al. 1998),
making
it
possible
to
determine
its
physical
parameters.
We
used
the
parameters
derived
for
the
Galactic
$N_H$
value
as
upper
limits
for
the
luminosity
and
radius
and a
lower
limit
for
$T_{\rm eff}$, and
used
the
parameters
derived
allowing
$N_H$
to
be
free
for
the
opposite
limits.
The
location
of
the
source
in
the
HR
diagram
is
near
the
turning
points
of
evolutionary
tracks
with
masses
of
1.1 - 1.2
$M_{\odot}$ for
hot
and
0.8 -0.9
$M_{\odot}$ for
cool
white
dwarfs.
Its
size
corresponds
to
the
theoretical
radius
of
white
dwarfs
with
masses
of
0.6 - 0.7
$M_{\odot}$, or
1 - 1.3
$M_{\odot}$
for
the
2R
case.
The
source
size
exceeds
the
size
of a
cool
white
dwarf,
and
its
mass
lies
in
the
range
0.8 - 1.2
$M_{\odot}$.

{\it RXJ0527.8-6954.} The X-ray
flux from this object
continually
decreased
in
the
1990s,
but
the
source
was
not
detected
by
the
Einstein
satellite
20
years
ago,
although
it
was
in
its
field
of
view
(Greiner et al. 1991, 1996).
It
thus
appears
that
the
source
is
recurrent
and
was
in a
declining
brightness
phase
during
the
ROSAT
observations.
We
used
the
parameters
obtained
for
the
Galactic
$N_H$,
since
the
approximation
data
obtained
when
$N_H$
was
allowed
to
float
are
very
uncertain.
The
source
occupies
the
position
of
cooling
curves
for
white
dwarfs
with
masses
of
1.2 - 1.4
$M_{\odot}$ in
the
HR
diagram.
Its
position
in
the
$\log~g$ ---
$R$
plane
is
in
better
agreement
with
the
relation
for a
cool
white
dwarf
with
$\log~g$
=9.5,
which
yields a
mass
based
on
its
size
of
1.3 - 1.4
$M_{\odot}$. Thus,
the
$\log~g$
of
the
source
is
fairly
high
(about
9.5)
and
its
mass
is
in
the
range
1.2 - 1.4
$M_{\odot}$.

{\it CAL
87.}
This
object
is
an
eclipsing
binary
system
that has been studied by various authors and observed
by
various
satellites
(Hartmann \& Heise 1997,
Parmar et al. 1997,
Asai
et
al. 1998).
The
$N_H$
value
is
appreciably
higher
than
the
Galactic
value,
presumably
due
to
the
fact
that
the
orbital
plane
is
strongly
inclined
to
the
line
of
sight,
and
the
X-ray
source
is
partially
eclipsed
by
material
above
the
plane
of
the
disk.
Its
parameters
are
well
defined,
since
it
has a
hard
spectrum.
The
position
of
the
object
in
the
HR
diagram
suggests
that
the
$N_H$
value
in
the
direction
of
the
source
may
be
overestimated,
so
that
its
luminosity
and
size
are
likewise
overestimated.
We
therefore
considered
only
the
lower
limits
of
these
quantities.
The source is located
in
the HR
diagram
in
the SBS
near white dwarfs with masses of
1.1 - 1.3
$M_{\odot}$. Its size
is
in
agreement
with
theoretical
radii
of
white
dwarfs
with
masses
lower
than
1.05
$M_{\odot}$ (1.3
$M_{\odot}$ for
the
2R
case).
Consequently,
the
size
of
the
source
exceeds
those of cool white dwarfs, and its mass is in the range
1.1 - 1.3
$M_{\odot}$.

{\it CAL
83.}
This
is a
classical
double
supersoft
source.
We
used
the
parameters
obtained
for
the
Galactic
$N_H$
value,
since
the
absorption
for
this
object
derived
from
HST
observations
is
close
to
this
value
(G\"ansicke et al. 1998).
The
source
is
located
in
the
HR
diagram
below
the
SBS
on
the
cooling
curves
for
white
dwarfs
with
masses
of
0.8 - 1.1
$M_{\odot}$ (for
$\log~g$
=8.0 - 9.5).
Its
position
in
the
$\log~g$ ---
$R$
plane
is
in
agreement
with
the
size
of a
cool
white
dwarf
with a
low
$\log~g$
(about
8.0).
The
mass
derived
from
the
size
corresponds
to
0.6 - 0.8
$M_{\odot}$ (1.2 - 1.3
$M_{\odot}$ for
the
2R
case).
Thus,
our
estimates
suggest
the
mass
is
close
to
0.8
$M_{\odot}$.

{\it 1E
0035.4 - 7230.}
This
is a
cataclysmic
variable
that
probably
has a
complex
history
(Kahabka \& Ergma 1997).
We
derived
parameters
for
two
separate
observations,
which
are
in
good
agreement
with
each
other.
We
used
the
parameters
obtained
when
approximating
the
April
28,
1992
observations
using
the
Galactic
$N_H$
value.
The source is located
in
the HR diagram
below
the
SBS
on
cooling
curves
for
white
dwarfs
with
masses
0.8 -1.0
$M_{\odot}$.
Its
size
is
in
agreement
with
that
for a
cool
white
dwarf
with a
mass
of
about
0.75
$M_{\odot}$
(1.25
$M_{\odot}$ in
the
2R
case).
Consequently,
the
size
and
position
of
the
source
in
the
HR
diagram
indicate
the
low
value
log =
8.0
and a
mass
of
0.7 - 0.8
$M_{\odot}$.

{\it RX
J0048.4-7332}.
This
is a
symbiotic
novae
that
has
been
in a
flaring
stage
since
1981
(Morgan 1992).
Our
hydrostatic
models
yield
high
luminosities
and
sizes
that
are
not
consistent
with
the
vdH92
model.
However,
models
with
winds ($\dot M \sim
10^{-5} - 10^{-6}$ $M_{\odot}$/yr)
yield
reasonable
estimates
of
the
source's
luminosity
(Jordan,
et al, 1996).
The
parameters
we
have
determined
should
be
treated
as
estimates.
We
are
not
able
to
determine
the
source's
mass.

{\it 1E
0056.8-7154.}
This
is
the
nucleus
of
the
planetary
nebula
N67.
We
obtained
parameters
for
two
observations
that
were
not
in
good
agreement,
although
the
confidence
contours
for
the
two
observations
are
consistent
with
each
other;
i.e.,
there
is a
high
probability
that
the
parameters
obtained
from
the
second
observation
are
also
possible
for
the
first
observation.
In
our
analysis,
we
used
the
parameters
derived
for
the
October
7,
1993
observation
leaving
$N_H$
free
to
vary with
$\log~g$
=8.0.The object is located in the SBS
for
white
dwarfs
with
masses
of
0.9 - 1.1
$M_{\odot}$ (hot
white
dwarfs)
and
0.5 - 0.8
$M_{\odot}$ (cool
white
dwarfs).

Its
size
exceeds
the
sizes
of
cool
white
dwarfs
and
is
consistent
with
the
2R
case
if
the
mass
is
lower
than
1.0
$M_{\odot}$. It is probable that the white dwarf is hot,
has
recently
formed
(the
planetary
nebula),
and
has a
mass
of
0.9 - 1.1
$M_{\odot}$.

{\it RX
J0019.8+2156
or
QR
And.}
This
is a
Galactic
supersoft
source.
We
used
the
parameters
obtained
leaving
the
$N_H$
value free with
$\log~g$
=7.5, since
the
parameters
derived
for
the
Galactic
$N_H$
value
yield a
very
low
luminosity
($ \sim
2 \cdot
10^{34}
erg/s d(kpc)^2$) and
size
(4 $\cdot
10^7$
cm $d(kpc)$)
for
the
source.
The
results
are
very
uncertain,
but
in
good
consistency
with
the
vdH92
model.
We
are
not
able
to
estimate
the
source
mass
using
the
available
data.

{\it RX
J0925.8-6809.}
This
is
also a
Galactic
supersoft
source
located
behind a
molecular
cloud
in
Vela,
so
that
it
is
strongly
reddened
($E(B-V)$=2.1)
(Motch et al. 1994).
We
adopted a
distance
of 2
kpc,
since
this
yields
good
agreement
with
the
vdH92
model.
In
this
case,
the
source
is
located
above
the
SBS,
and
its
mass
is
1.2 - 1.4
$M_{\odot}$.
Its
size
is
also
consistent
with
this
mass
estimate
(larger
than
0.95
$M_{\odot}$ for a
cool
white
dwarf
and
larger
than
1.25
$M_{\odot}$ for
the
2R
case).
The
size
of
the
white
dwarf
is
larger
than
expected
for a
cool
white
dwarf.

Thus,
we
have
estimated
the
white-dwarf
masses
and
the
most
probable
values
of
the
remaining
physical
parameters
for
seven
supersoft
sources
(Table
2).
Of
these,
four
sources
(CAL
87,
RX
J0925,
1E0056,
and
RX
J0513)
were
in
the
SBS
at
the
epoch
of
observation,
with
their
sizes
exceeding
those
of
cool
white
dwarfs.
The
remaining
three
sources
(CAL
83,
RX
J0527,
and
1E0035)
were
below
the
SBS
on
white-dwarf
cooling
curves,
with
their
sizes
consistent
with
those
of
cool
white
dwarfs.

We
conclude
that
the
physical
characteristics
of
classical
supersoft
sources
are
consistent
with
the
model
of
stable
or
recurrent
thermonuclear
burning
on
the
surface
of a
white
dwarf
and
that
these
sources
occupy
positions
in
the
HR
diagram
either
near
the
turning
points
of
evolutionary
tracks
for
this
model
or
lower,
on
white-dwarf
cooling
curves.
\bigskip

\centerline{\bf 4. Mass --- temperature relation}
\bigskip

Let
us
investigate
the
relation
between
the
effective
temperature
of
the
supersoft
sources
and
the
corresponding
white-dwarf
masses.
It
is
known
(Iben \& Tutukov 1996)
that
the
luminosity
of a
white
dwarf
with
thermonuclear
burning
on
its
surface
is
located
on
the
horizontal
section,
or
plateau,
of
the
theoretical
curves
in
the
HR
diagram
associated
with
their
mass
(formulas
(2)
and
(3)).
Using a
linear
approximation
for
the
mass ---
radius
dependence
for
white
dwarfs
and
relation
(3)
between
the
luminosities
and
masses
of
white
dwarfs,
Iben
and
Tutukov
(Iben \& Tutukov 1996)
derived
the
dependence
of
the
maximum
possible
effective
temperature
$T_{\rm eff}$
on
the
white-dwarf
mass:
\be
T_{\rm eff} = 3.6 \cdot 10^5 K
\frac{(M/M_{\odot} - 0.26)^{1/4}}{(1-0.59 M/M_{\odot})^{1/2}},
\ee
shown
in
Fig. 4
by
the
dotted
curve.
This
formula
was
derived
assuming
that
the
maximum
temperature
is
reached
when
the
luminosity
is
half
the
luminosity
on
the
plateau
(3),
and
the
photospheric
radius
is
double
the
radius
of
the
white
dwarf.

This
same
figure
presents
the
positions
of
the
four
sources
located
in
the
SBS
and
two
theoretical
dependences.
The
upper
curve
is
constructed
from
the
positions
of
the
maximum-temperature
points
for
the
evolutionary
tracks
of
novae
in
the
declining
brightness
phase
(Kato \& Hachisu 1994),
and
the
lower
curve
from
the
positions
of
these
points
for
hot
white
dwarfs
with a
shell
burning
source
(Iben \& Tutukov 1996).
It
is
clear
that
the
formula
proposed
in
(Iben \& Tutukov 1996)
does
not
provide a
good
description
of
the
relation between the mass and maximum temperature
of
the
supersoft
sources,
although
its
underlying
principle
remains
correct.
We
attempted
to
improve
this
formula
by
allowing
for
the
fact
that
the
mass --- radius
relation
for
white
dwarfs
is
not
linear,
and
the
ratio
of
the
size
of
the
photosphere
of a
supersoft
source
to
the
radius
of
the
corresponding
cool
white
dwarf
could
depend
on
its
mass.
We
accordingly
approximated
the
mass --- radius
relation
using a
third-order
polynomial:
\be
\frac{R}{R_{\odot}}=0.0273 - 0.0417
\frac{M}{M_{\odot}} +
0.0364
\left(\frac{M}{M_{\odot}}\right)^2 - 0.0138
\left(\frac{M}{M_{\odot}}\right)^3,
\ee
and
used
this
in
place
of
the
linear
relation.
However,
Accordingly,
we
propose
the
following
formula
for
this
significantly
influenced
the results only for
white
the
relation
between
the
mass
of a
white
dwarf
and
its
dwarfs
with
masses
higher
than
1.2
$M_{\odot}$.
Analysis of
the
ratio
of
the
photospheric
radius
to
the
white-
dwarf
radius
indicated
that,
in
order
to
construct
the
best
approximation
to
the
results
of
the
computations
of
surface
thermonuclear
burning,
this
ratio
should
be
inversely
proportional
to
the
square
of
the
white
dwarf
mass.

Analysis
of
effective
temperature
in
the
region
of
the
SBS:
\be
T_{\rm eff} = 3 \times 10^5 K \frac{M/M_{\odot}(M/M_{\odot} - 0.26)^{0.25}}
{(1 - 1.53M/M_{\odot} +1.33(M/M_{\odot})^2 - 0.51(M/M_{\odot})^3)^{0.5}}.
\ee
This
relation
is
shown
in
Fig.
4a
by
the
dashed
curve.
We
can
see
that
it
provides a
good
approximation
to
the
results
of
the
numerical
computations.
It
was
obtained
assuming
that
the
luminosity
of
the
white
dwarf
is
half
the
luminosity
on
the
plateau,
and
the
photospheric
radius
is a
factor
of
$2(M_{\odot}/M)^2$
larger
than
the
white-dwarf
radius.
We
note
that
this
is
only a
suggestion
for
the
required
relation,
and
the
actual
relation
at
the
boundary
of
the
SBS
could
be
more
complex.

Figure
4b
shows
the
positions
of
the
remaining
three
sources
(CAL
83,
1E
0035,
and
RX
J0527)
in
the
$M$ --- $T_{\rm eff}$
plane.
The
solid
curves
depict
the
dependences
of
the
maximum
temperature
on
the
white-
dwarf
mass
that
follow
from
the
numerical
computations
(the
same
as
in
Fig.
4a).
The
dashed
curves
represent
the
relationships
between
the
effective
temperatures
and
masses
of
white
dwarfs
for
sources
with
radii
equal
to
the
radii
of
cool
white
dwarfs
and
luminosities
that
are
factors
of 4
to
40
lower
than
the
maximum
possible
luminosity
(3).
The
numbers
near
the
curves
indicate
the
ratio
of
the
luminosity
of
the
white
dwarf
to
the
maximum
luminosity.
The
degree
of
cooling
of
these
three
sources
at
the
observation
epoch
and
the
influence
of
the
nonlinearity
of
the
mass --- radius
relation
for
white
dwarfs
are
evident
(compare
the
dotted
curve
in
Fig.
4a
and
the
dashed
curves
in
Fig.
4b).
\bigskip

\centerline{\bf 5. Conclusion}
\bigskip

We
have
analyzed
archival
ROSAT
observations
of
10
known
supersoft
X-ray
sources,
and
investigated
the
source
parameters
obtained
in
the
first
part
of
our
study
(Ibragimov et al. 2003)
to
test
for
consistency
with
the
model
of
thermonuclear
hydrogen
burning
on
the
surface
of a
white
dwarf
(vdH92).
The
values
of
$T_{\rm eff}$
and
$\log~g$
obtained
for
the
three
hottest
sources
are
consistent
with
those
expected
for
the
stable-burning
strip
in
this
model.
The
parameters
of
the
remaining
sources
are
consistent
with
those
expected
for
this
strip
if
the
lowest
possible
values
of
$\log~g$ (7.5 - 8.0)
are
adopted.

We
have
estimated
the
sizes
and
luminosities
of
ten sources. The sizes and luminosities of two sources
(RX
J0048
and
RX
J0439)
are
not
consistent
with
those
for
hot
white
dwarfs,
confirming
earlier
conclusions
of
other
studies
that
hydrostatic
white-dwarf
model
atmospheres
with
solar
chemical
composition
are
not
suitable
for
these
sources.
The
parameters
of
another
source
(RX
J0019)
are
very
uncertain,
although
they
lie
in
the
range
of
admissible
parameters
for
the
hot
white-dwarf
model.

The
sizes
and
luminosities
of
seven
classical
super soft sources are in good agreement with the vdH92
model, enablingus to estimate their masses from their
sizes and positions in the HR diagram. We considered
the
dependence
of
$T_{\rm eff}$
on
the
mass
of
the
white
dwarfs
in
these
supersoft
sources
and
have
proposed a
formula
that
can
be
used
to
estimate
the
mass
of a
white
dwarf
in a
classical
binary
supersoft
source
from
its
temperature.
\bigskip

\centerline{\bf Acknowledgments}
\bigskip

This
work
was
supported
by
the
Russian
Foundation
for
Basic
Research
(project
nos.
99-02-17488
and
02-02-17174).

\newpage

\centerline{\bf References}
\bigskip

K.
Asai,
T.
Dotani,
F.
Nagase,
et
al., 1998,
Astrophys.
J.
503,
L143

A.
Cowley,
P.
Schmidtke,
D.
Crampton,
and
J.
Hutchings, 1998,
Astrophys.
J.
504,
854

B.
G\"ansicke,
A.
van
Teeseling,
K.
Beuermann,
\&
D.
De
Martino, 1998,
Astron.
Astrophys.
333,
163

B.
G\"ansicke,
A.
van
Teeseling,
K.
Beuermann,
\&
K.
Reinsch, 2000,
New
Astron.
Rev.
44,
143

J.
Greiner,
G.
Hasinger, \&
P.
Kahabka, 1991,
Astron.
Astrophys.
246,
L17

J.
Greiner,
R.
Schwarz,
G.
Hasinger,
\&
M.
Orio, 1996,
Astron.
Astrophys.
312,
88

M.
Fujimoto, 1982,
Astrophys.
J.
257,
767

H.
Hartmann
\&
J.
Heise, 1997,
Astron.
Astrophys.
322,
591

I.
Iben, 1982,
Astrophys.
J.
259,
244

I.
Iben
\&
A.
V.
Tutukov, 1993,
Astrophys.
J.
418,
343

I.
Iben
and
A.
V.
Tutukov, 1996,
Astrophys.
J.,
Suppl.
Ser.
105,
145

A.
A.
Ibragimov,
V.
F.
Suleimanov,
A.
Vikhlinin,
\&
N.A.Sakhibullin, 2003,
Astron.
Rep.
47,
186

S.
Jordan,
W.
Schmutz,
B.
Wolff,
et
al., 1996, Astron.
Astrophys.
346,
897

P.
Kahabka
\&
E.
Ergma, 1997,
Astron.
Astrophys.
318,
108

P.
Kahabka,
H.
Hartmann,
A.
Parmar,
\&
I.
Negueruela, 1999,
Astron.
Astrophys.
347,
L43

P. Kahabka \&
E.
P.
J.
van
den
Heuvel, 1997,
Ann.
Rev.
Astron.
Astrophys.
35,
69

M.
Kato
and
I.
Hachisu, 1994,
Astrophys.
J.
437,
802

D.
Morgan, 1992, MNRAS,
258,
639

C.
Motch,
G.
Hasinger,
\&
W.
Pietsch, 1994,
Astron.
Astrophys.
284,
827

K.
Nomoto, 1982,
Astrophys.
J.
253,
798

B.
Paczynski
\&
A.
Zytkow, 1978,
Astrophys.
J.
222,
604

A.
Parmar,
P.
Kahabka,
H.
Hartmann,
et
al., 1997, Astron.
Astrophys.
323,
L33

R.
Popham
\&
R.
Narayan, 1995,
Astrophys.
J.
442,
337

E.
P.
J.
van
den
Heuvel,
D.
Bhattacharya,
K.
Nomoto,
and
S.
Rappaport, 1992,
Astron.
Astrophys.
262,
97

\begin{table}
\caption{
Approximation
parameters}\label{datas}
\begin{center}
\begin{tabular}{l|c|c|c|c|c|c}
\hline
 & & & & & & \\
Object &  $N_H$,   & $T_{\rm eff}$,   & $\log~g$
& $\log~(R/d)^2$      & Flux &  $\chi^2/$  \\
&             $10^{20}$ cm$^{-2}$ & $10^5$ K      &  $cm^2 s^{-1}$ &
& 0.2-2 keV &   d.o.f.              \\
 & & & & & erg/cm$^{-2}$/s & \\
 & & & & & & \\
\hline
 & & & & & & \\
RX J0439.8-6809 & 5.60 &$ 2.72_{-0.15}^{+0.12}$& 7.5 &
-26.24$^{+0.35}_{-0.26}$  &$7.99\cdot 10^{-11}$&  9.30/9 \\
 & & & & & & \\
RX J0513.9-6951 & 7.24 &$ 5.74_{-0.03}^{+0.03}$& 8.4 &
-28.29$^{+0.02}_{-0.01}$ &$1.71\cdot 10^{-10}$&  33.8/9 \\
 & & & & & & \\
RX J0527.8-6954 & 6.31 & $5.60_{-1.68}^{+0.37}$ & 9.5 &
-29.59$^{+0.95}_{-0.14}$  &  $6.84\cdot 10^{-12}$&2.01/9 \\
 & & & & & & \\
CAL 87   & $69.2_{-22.0}^{+23.1}  $ & $8.20_{-0.24}^{+0.22}$ & 9.0 &
-28.20$^{+0.85}_{-0.81}$ & $1.17\cdot 10^{-9} $ & 13.5/8 \\
 & & & & & & \\
CAL 83   & 6.33 & $5.04_{-0.17}^{+0.17}$ & 8.0 & -28.57$^{+0.08}_{-0.08}$
 & $4.59\cdot 10^{-11}$&17.3/9 \\
 & & & & & & \\
1E 0035.4-7230  & 6.94 &$ 4.64_{-0.13}^{+0.09} $& 8.0 &
-28.56$^{+0.08}_{-0.05}$ &$2.83\cdot 10^{-11}$&  6.10/9 \\
 & & & & & & \\
RX J0048.4-7332 & $27.6_{-12.6}^{+10.5}  $ & $3.35_{-0.48}^{+0.55}$ & 7.5 &
-24.55$^{+2.82}_{-2.50}$ & $3.09 \cdot 10^{-8} $ &
8.08/8 \\
 & & & & & & \\
1E 0056.8-7154  & $3.71_{-2.38}^{+21.29}$ & $4.00_{-0.97}^{+1.24}$ & 8.0  &
-28.85$^{+2.02}_{-0.44}$ &  $5.31\cdot 10^{-12}$& 13.2/8\\
 & & & & & & \\
RX J0019.8+2156 & $16.7_{-13.35}^{+6.9} $ & $2.80_{-0.35}^{+1.26} $ & 7.5 &
-23.62$^{+2.82}_{-4.67}$ & $ 4.46\cdot 10^{-8}$&18.1/8  \\
 & & & & & & \\
RX J0925.7-4758 & $163_{-46}^{+14}      $ & $9.85_{-0.37}^{+1.05} $ & 9.5 &
-26.23$^{+0.59}_{-1.57}$ &$2.53 \cdot 10^{-7}$ & 8.85/8 \\
 & & & & & & \\
\hline
\\
\multicolumn{7}{l}{
Note:
The
lack
of
an
error
for
$N_H$
means
that
this
parameter
was
fixed
at
the
Galactic
value.}\\
\multicolumn{7}{l}{
The
fluxes
were
corrected
for
interstellar
absorption.}\\
\end{tabular}
\end{center}
\end{table}

\newpage
\begin{table}
\caption{
Estimates
of
physical
parameters
of
the
sources}

\label{data_cool}
\begin{center}
\begin{tabular}{l|c|c|c|c|c} \hline
 & & & & &  \\
Source & $T_{\rm eff}$,~$10^5$ K & $\log~g$ & $R, 10^8$ cm & $M/M_{\odot}$
& $L_{\mathrm{bol}}$, $10^{37}$ erg s$^{-1}$\\
 & & & & &    \\
\hline
 & & & & &      \\
RX J0513 & 5.85$\pm$0.1       & 8.0-8.5 & 8-12 & 0.8-1.2 & 5.5-9.5\\
 & & & & &      \\
RX J0527 & 5.60$^{+0.37}_{-1.68}$  & 9.0-9.5 & 2.5-4 & 1.2-1.4 & 0.4-0.9\\
 & & & & &      \\
CAL 87 & 8.2$\pm$ 0.25         & 8.5-8.75 & 5-9 & 1.1-1.3 & 7.5 - 20\\
 & & & & &      \\
CAL 83 & 5.04$\pm$ 0.17        & 8.0-8.25 & 7-9 & 0.7-0.9 & 2.8-3.1\\
 & & & & &      \\
1E 0035 & 4.64$\pm$ 0.1        & 7.7-7.9 & 8-10 & 0.65-0.85 & 3-3.3\\
 & & & & &      \\
1E 0056 & 4$\pm$1              & 7.7-8.5 & 10-28 & 0.85-1.05& 2.4 - 8 \\
 & & & & &      \\
RX J0925 & 9.85$^{+1}_{-0.4}$  & 8.5-9.5 & 5-9 & 1.2-1.4 & 4-20\\
 & & & & &      \\
\hline \end{tabular} \end{center} \end{table}

\clearpage
\newpage

\centerline{\bf Captions to figures}
\bigskip

{\bf Figure 1.}
Position
of
the
sources
in
the
Hertzsprung - Russell
diagram.
Theoretical
relations
for
white
dwarfs
with
hydrogen
burning
on
their
surfaces
are
also
shown:
(a)
evolutionary
tracks
for
novae
in
the
declining
brightness
phase
(Kato \& Hachisu 1994);
(b)
theoretical
curves
for
stable
burning
of
hydrogen
and
helium
on
the
surfaces
of
hot
white
dwarfs
(Iben \& Tutukov 1996).
The
bold
solid
curve
shows
the
boundary
of
the
stable-burning
strip.
The
numbers
near
the
curves
indicate
the
white-dwarf
masses
in
solar
units.
\bigskip

{\bf Figure 2.}
Position
of
the
sources
in
the
$R$ ---
$\log~g$
plane:
(a)
with
specified
values
of
$\log~g$
or
sizes
that
are
not
consistent
with
those
of
white
dwarfs;
(b)
with
uncertain
values
of
$\log~g$.
The
solid
curve
shows
the
theoretical
dependence
for
white
dwarfs
that
follows
from
the
mass --- radius
relation,
and
the
dashed
curve
shows
the
same
dependence
for
the
case
when
the
white-dwarf
radius
is
doubled.
\bigskip

{\bf Figure 3.}
Theoretical
$M$ ---
$\log~g$
for
cool
white
dwarfs
(solid)
and
for
objects
with
the
same
masses
but
doubled
radii.
\bigskip

{\bf Figure 4.}
Position
of
the
studied
sources
in
the
$T_{\rm eff}$ ---
$M$
plane.
(a)
Position
of
sources
in
the
stable-burning
strip.
The
dotted
curve
labeled
IT96
corresponds
to
the
theoretical
relation
proposed
by
Iben
and
Tutukov
(1996) .
The
solid
curves
show
relations
between
the
maximum
temperatures
and
masses
of
white
dwarfs
following
from
numerical
computations
of
brightness
declines
of
novae
(Kato \& Hachisu 1994)
(upper
curve,
compare
with
Fig.
1a)
and
of
stable
hydrogen
and
helium
burning
on
the
surfaces
of
hot
white
dwarfs
(Iben \& Tutukov 1996)
(lower
curve,
compare
with
Fig.
1b).
The
dashed
curve
depicts
the
relation
of
(6),
which
we
proposed
to
estimate
the
masses
of
white
dwarfs
with
supersoft
sources
located
in
the
stable-burning
strip.
(b)
Position
of
sources
located
on
white-
dwarf
cooling
curves.
The
dashed
curves
show
the
dependence
of
the
source
temperature
on
the
white-dwarf
mass
if
its
radius
is
equal
to
the
radius
of a
cool
white
dwarf
with
this
mass
and
its
luminosity
comprises a
constant
fraction
of
the
luminosity
on
the
plateau
(3).
The
numbers
near
the
curves
indicate
the
luminosity
as a
fraction
of
the
maximum
luminosity
on
the
plateau.
\clearpage

\begin{figure}
\includegraphics[width=\columnwidth, bb=14 50 581 828, clip]{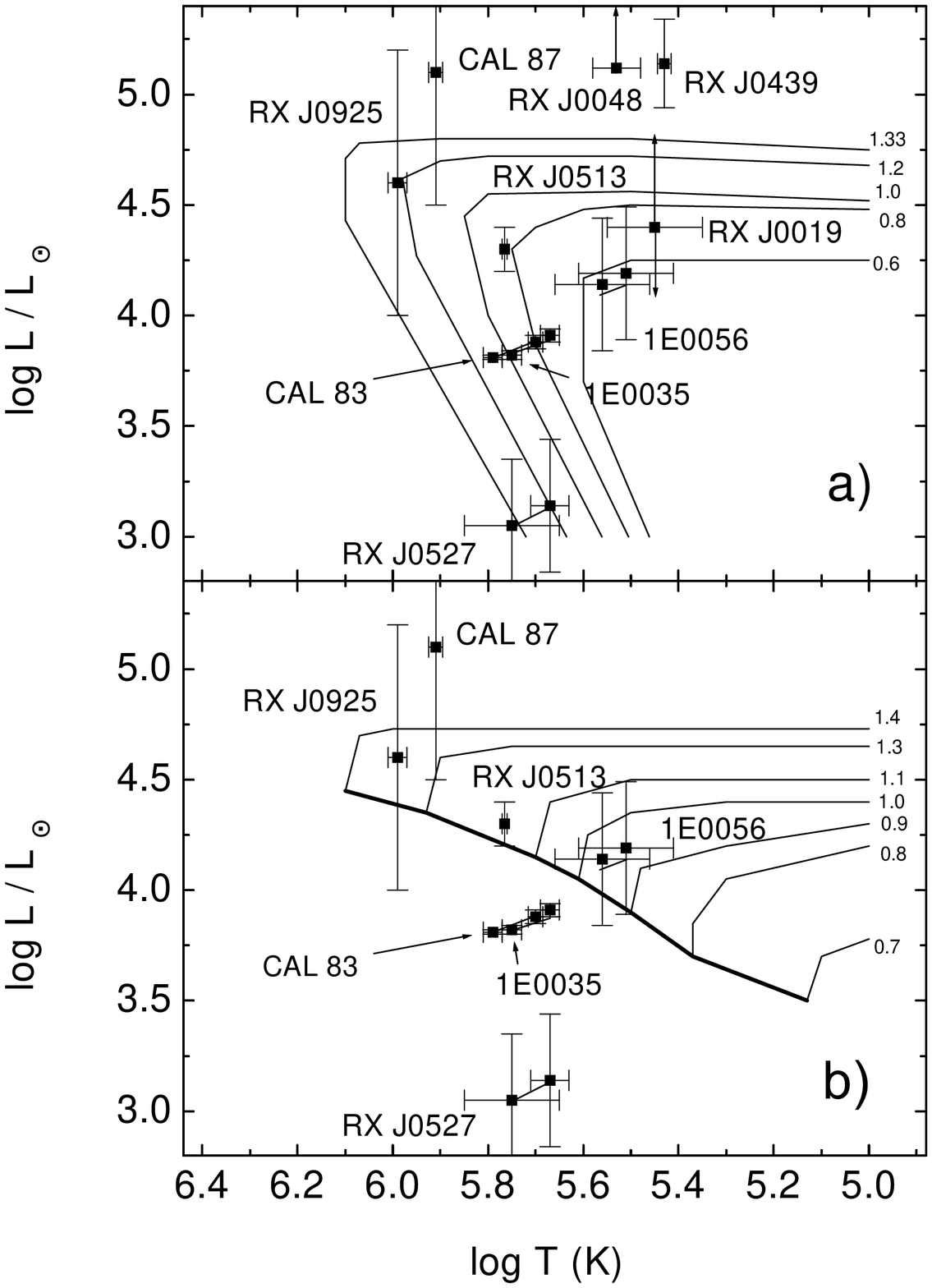}
\caption{\label{fig1}
}
\end{figure}

\begin{figure}
\includegraphics[width=\columnwidth, bb=14 50 581 828, clip]{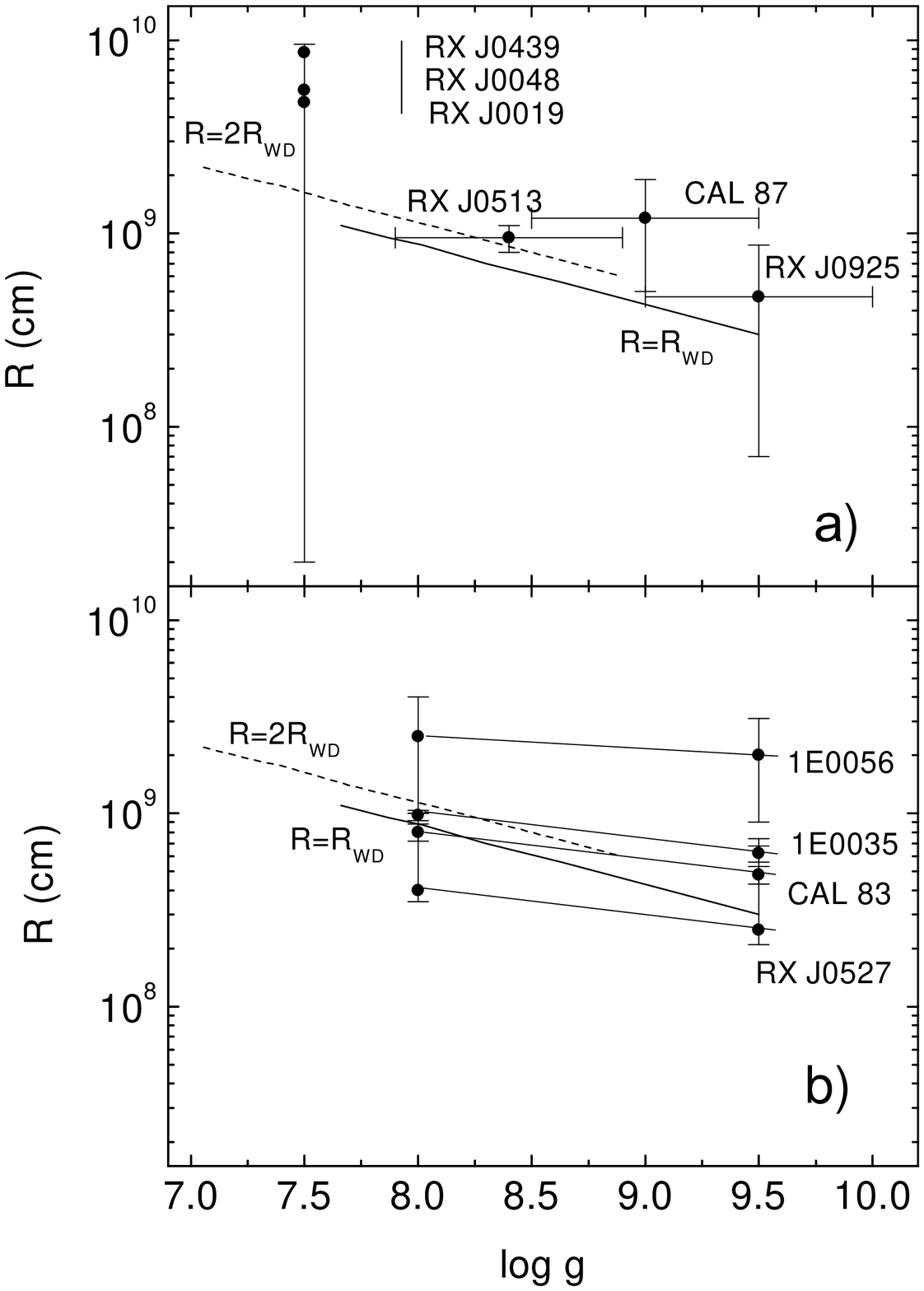}
\caption{\label{fig2}
}
\end{figure}

\begin{figure}
\includegraphics[width=\columnwidth, bb=14 50 581 828, clip]{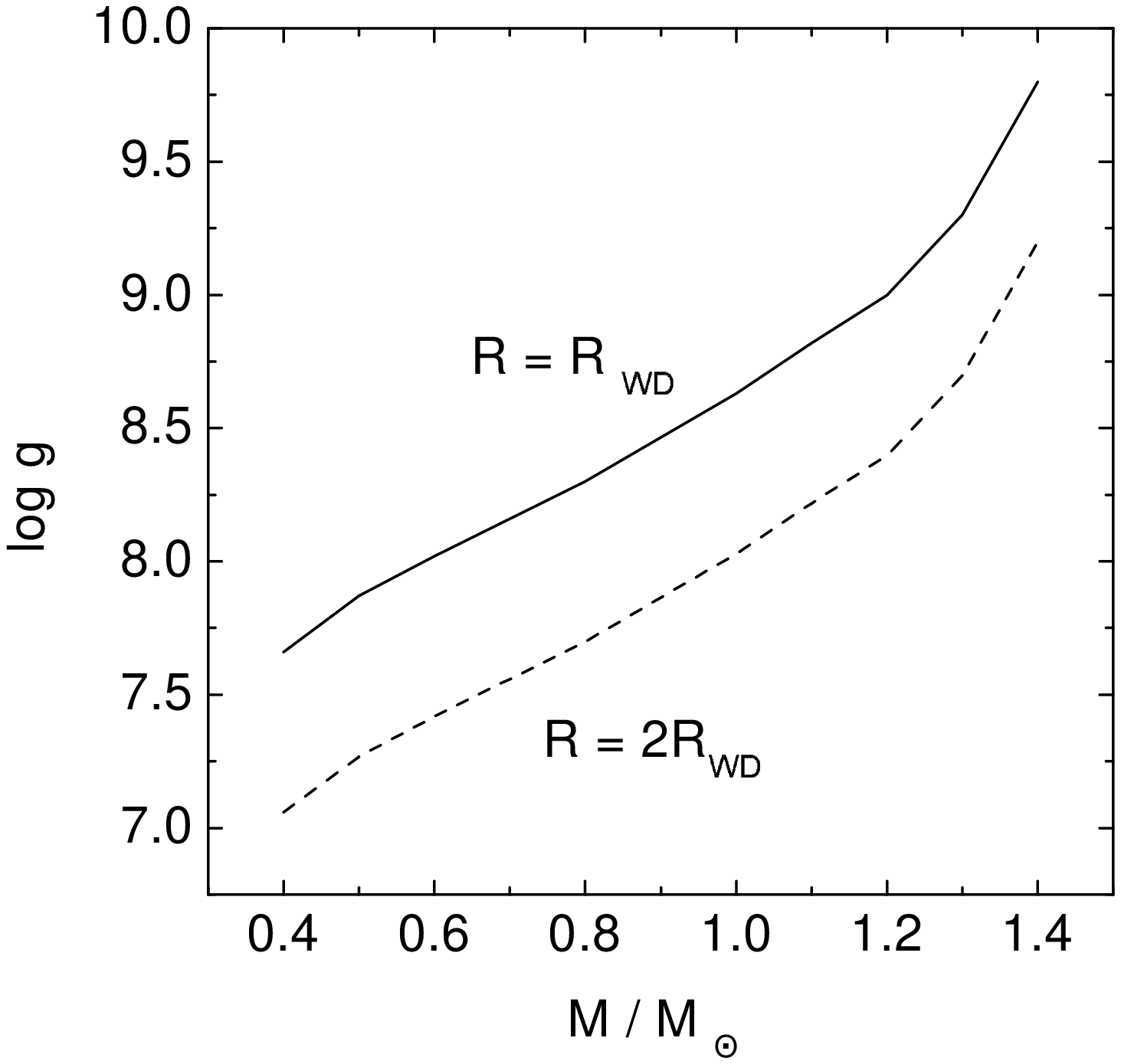}
\caption{\label{fig3}
}
\end{figure}

\begin{figure}
\includegraphics[width=\columnwidth, bb=14 50 581 828, clip]{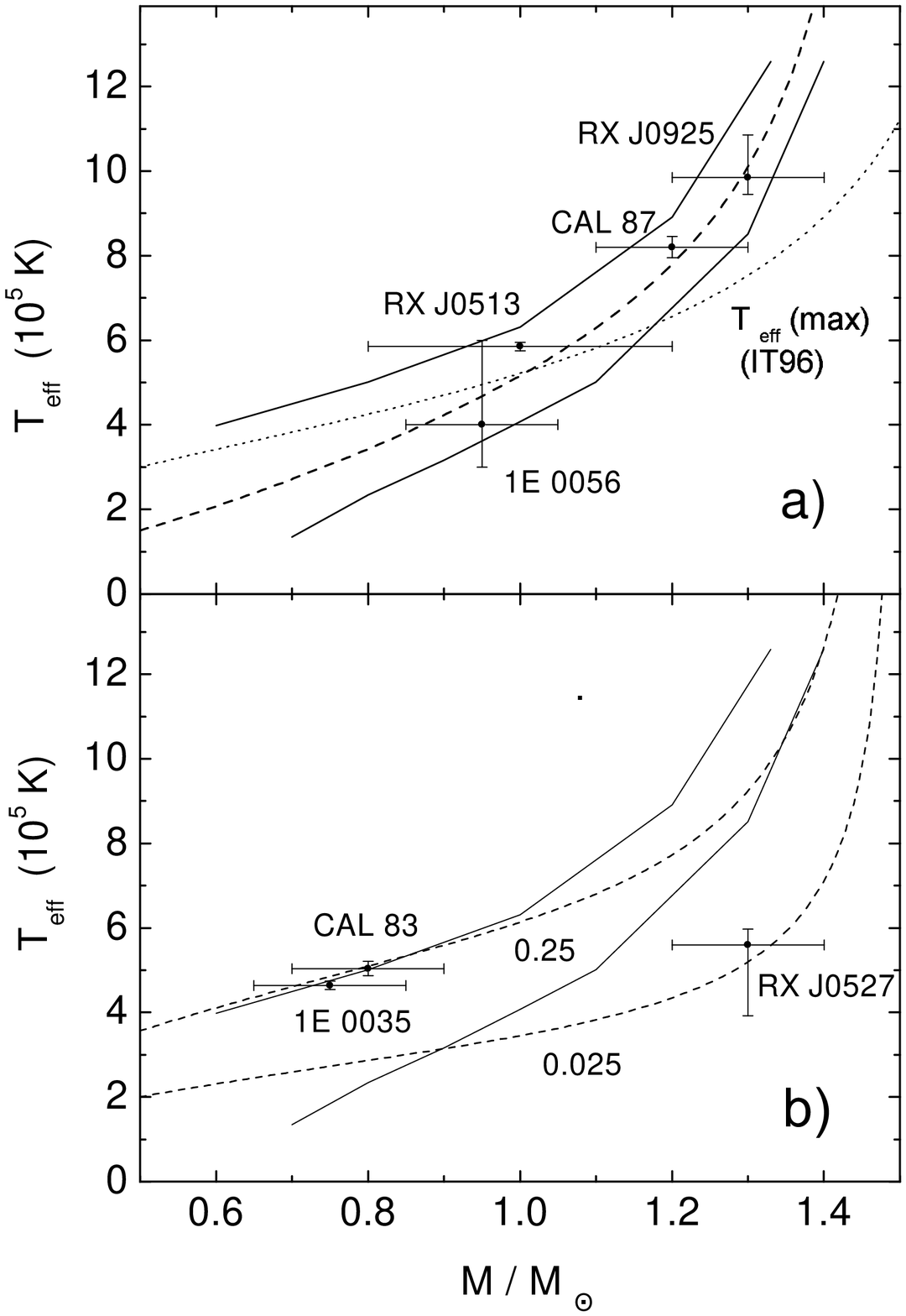}
\caption{\label{fig4}
}
\end{figure}
\end{document}